\documentclass[11pt,a4paper]{article}
\usepackage{graphicx}
\usepackage{caption} 
\usepackage{amsmath}
\usepackage{amsfonts}
\usepackage{amssymb}
\usepackage[square,numbers]{natbib}

\usepackage{url}
\usepackage{dcolumn}
\usepackage[margin=0.5in]{geometry}
\usepackage{tabu}
\newcommand{\w}{\omega}

\newcommand{\ba}{\begin{eqnarray}}
\newcommand{\ea}{\end{eqnarray}}
\newcommand{\be}{\begin{eqnarray}}
\newcommand{\ee}{\end{eqnarray}}

\begin{document}

\title{Storage Rings and Gravitational Waves: \\
Summary and Outlook}

\vspace{5mm}

\author{A.~Berlin$^1$,
M.~Br\"{u}ggen$^2$,
O.~Buchmueller$^{3}$, 
P.~Chen$^{4}$,  
R.~T.~D'Agnolo$^{5}$, 
R.~Deng$^{6}$,  \protect\\
J.~R.~Ellis$^{7,\times,}$\thanks{John.Ellis@cern.ch},
S.~Ellis$^{5}$,
G.~Franchetti$^{8}$, 
A.~Ivanov$^{9}$, 
J.~M.~Jowett$^{8}$, \protect\\
A.~P.~Kobushkin$^{10}$,
S.~Y.~Lee$^{11}$,
J.~Liske$^2$,
K.~Oide$^{12}$,
S.~Rao$^2$, 
J.~Wenninger$^{13}$, \protect\\
M.~Wellenzohn$^9$,
M.~Zanetti$^{14}$, 
F.~Zimmermann$^{13,\times,}$\thanks{Frank.Zimmermann@cern.ch}
\protect\\
\protect\\
$^\times${\it Editors}
\protect\\
\vspace*{2 mm} 
\protect\\ 
\small{$^1$New York U., USA;
$^2$U.~Hamburg, Germany; 
$^{3}$Imperial College London, UK;} \protect\\
\small{$^4$National Taiwan University, Taiwan;   
$^5$IPhT Paris, France;
$^6$SARI Shanghai, China;}
\protect\\ 
\small{$^{7}$King's College London, UK; 
$^8$GSI, Germany;
$^9$Vienna U., Austria;
$^{10}$BITP, Ukraine;}  \protect\\
\small{$^{11}$Indiana U., USA; 
$^{12}$KEK, Japan;
$^{13}$CERN, Switzerland; 
$^{14}$U.\& INFN Padua, Italy}
}

\maketitle
\begin{abstract} 
We report some highlights 
from the ARIES APEC workshop on ``Storage Rings and 
Gravitational Waves'' (SRGW2021),  held in virtual space 
from 2 February to 18 March 2021,  
and sketch a tentative landscape for using accelerators and 
associated technologies for the detection or generation
of gravitational waves.\\
\begin{center}
KCL-PH-TH/2021-28, CERN-TH-2021-068
\end{center}
\end{abstract}

\section{Introduction}
SRGW2021 \cite{srgw21} brought together storage-ring experts,  accelerator scientists, experimental particle physicists, theoretical physicists, astrophysicists, and members of the gravitational physics community. It surveyed and discussed 
various topics including: (1) examples of the sensitivities of storage rings to tides, earthquakes and other seismic perturbations; (2) earlier and more recent work on the use of storage rings to detect gravitational waves; (3) the dependence of the storage ring detection sensitivity on the ring size and other parameters; (4) possible enhancement strategies such a special optics, particular magnets and advanced diagnostics; (5)
gravitational radiation from storage ring beams;
(6) possible astrophysical and cosmological signals, their frequencies and amplitudes; (7) gravity research based on accelerator technologies; and (8) a tentative roadmap for
possible future R\&D and measurements. 

\begin{figure}[htbp]
   \centering
\includegraphics[width=0.49\columnwidth]{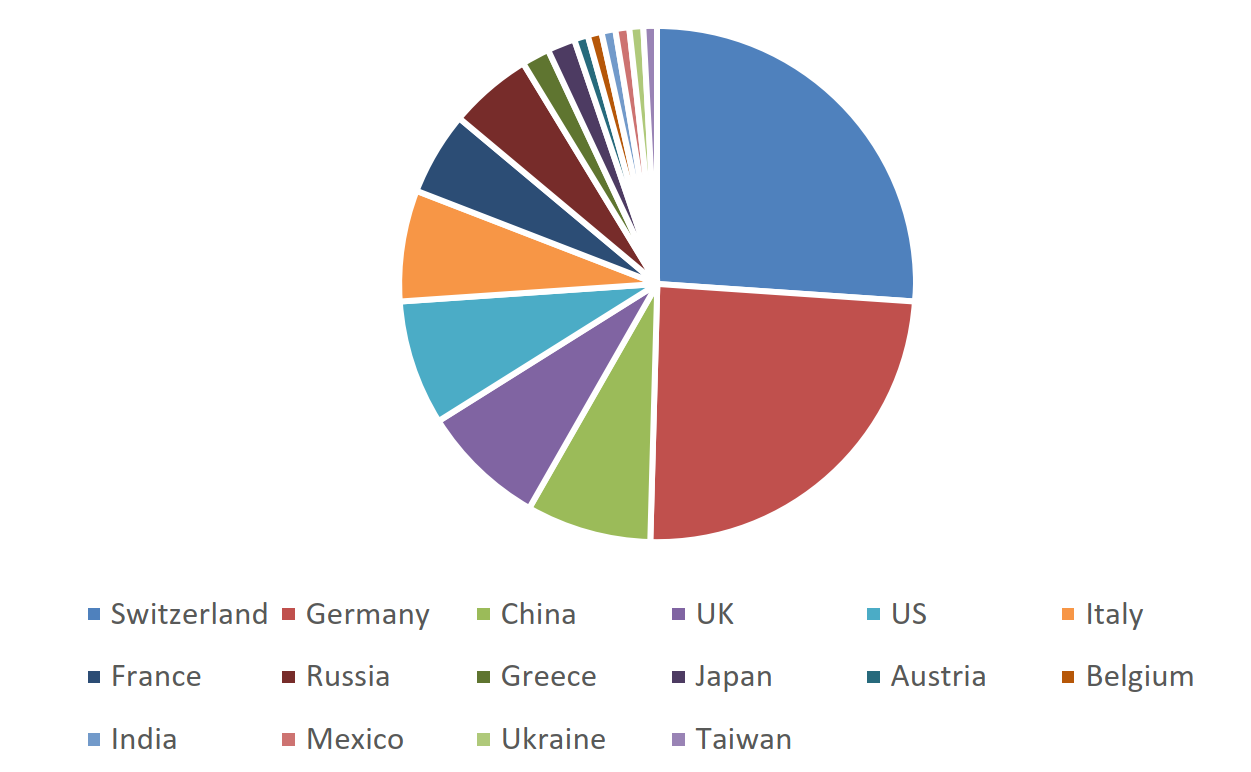}
\includegraphics[width=0.49\columnwidth]{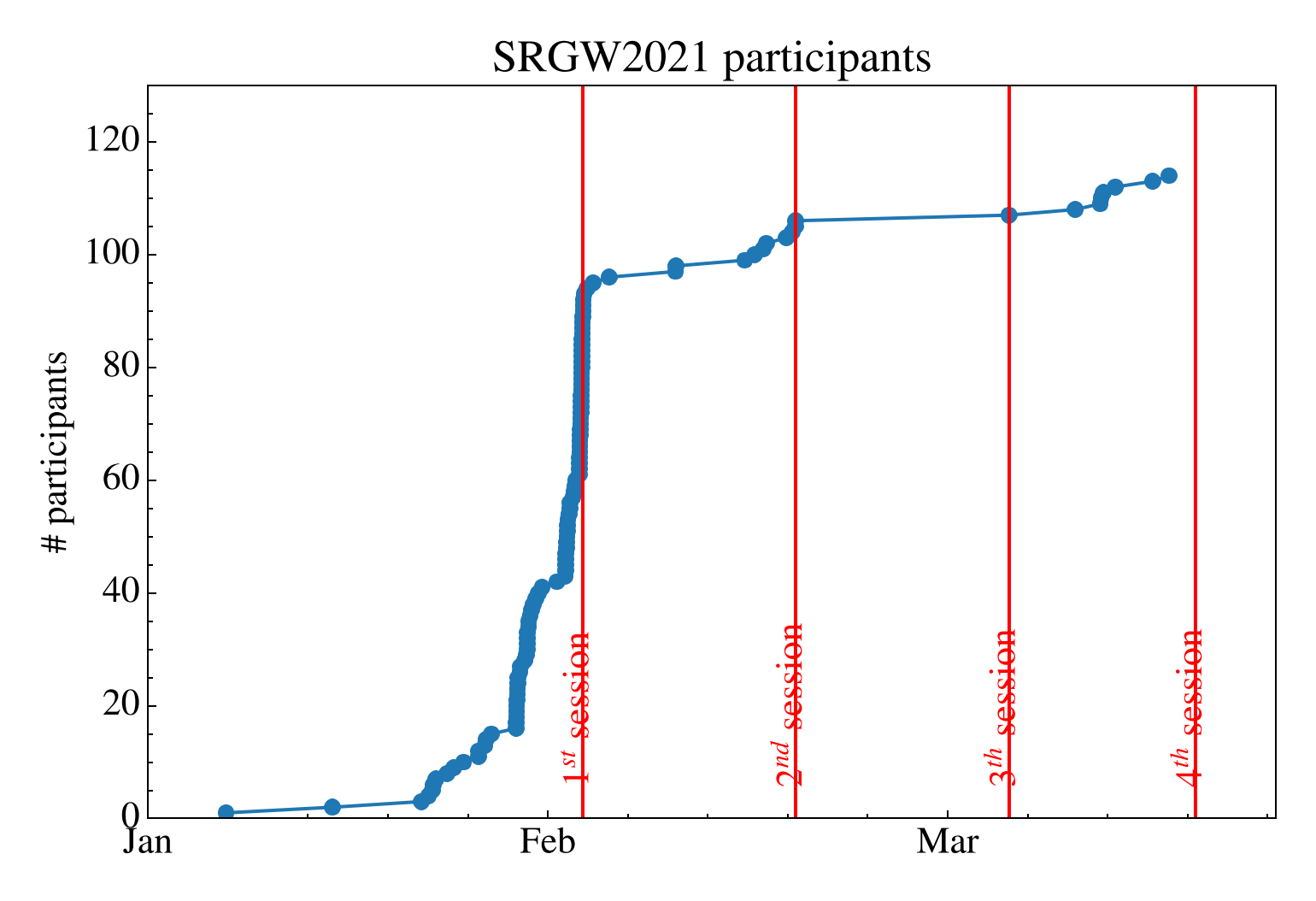}
   \caption{Geographical distribution of SRGW2021 participants 
   (left) and time evolution of registrations (right).
   \label{fig:parts}}
\end{figure}

SRGW2021  was held in 5 virtual sessions scheduled between 
2 February to 18 March 2021. 
A total of 115 participants registered for the event.
Figure \ref{fig:parts} shows the geographical distribution of 
the participants, along with the time evolution of their registrations. 
The workshop schedule, including its sessions, talks and speakers,
is shown in the Appendix.

\newpage

\section{Selected Highlights - Summaries by Presenters}

\subsection{Storage ring sensitivity to tides \& large-scale perturbations, earthquakes, noise - examples from LEP and LHC\\
{\it J.~Wenninger}}

\begin{figure}[hbp]
\centering
\includegraphics[width=0.6\textwidth]{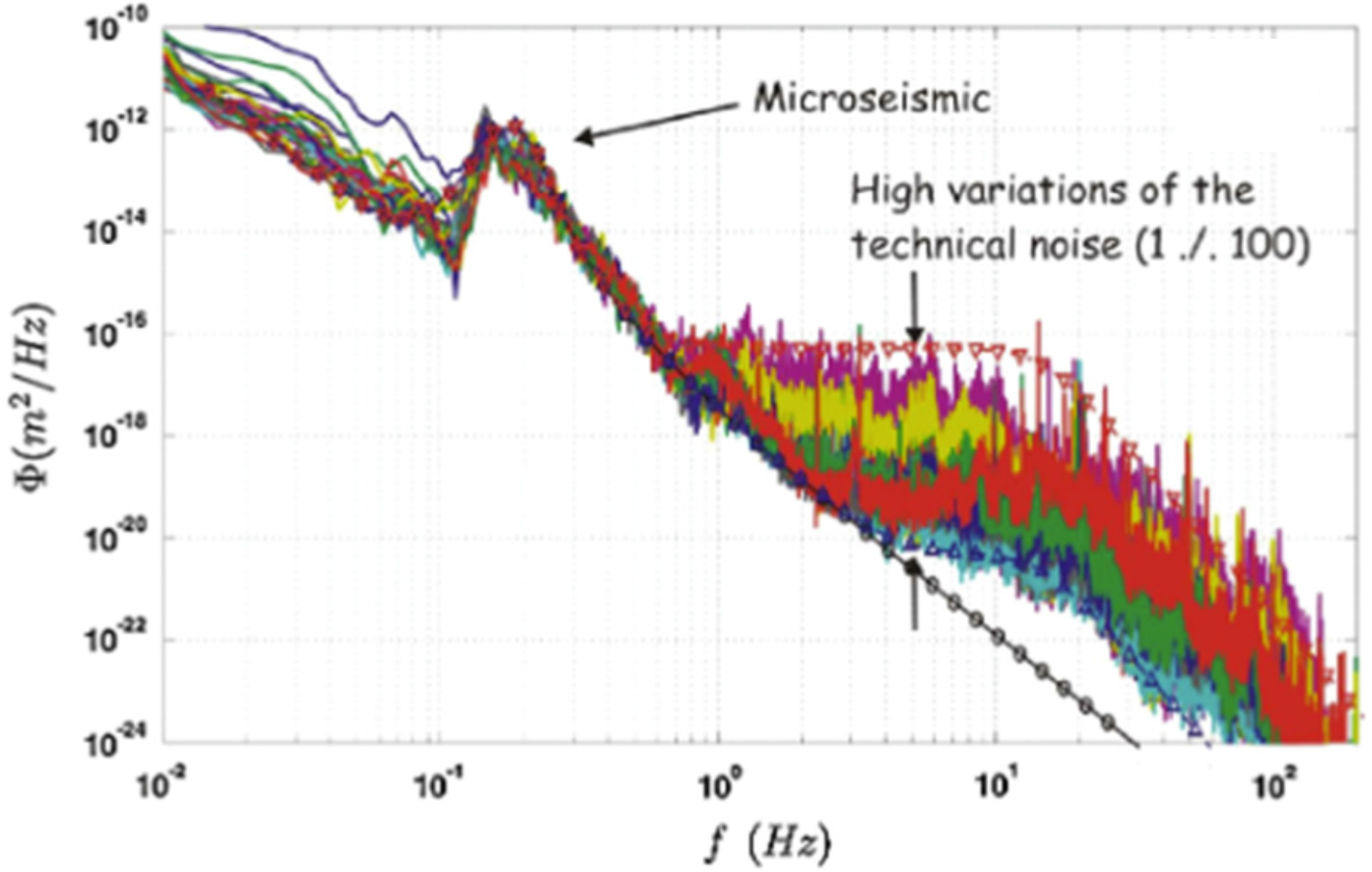}
\caption{Measurements of the seismic noise power spectrum in the LHC tunnel~\cite{CDGC}.}
\label{LHCSN}
\end{figure}

Tides and other large scale geological deformations like slow ground motion and earthquakes have been observed routinely with the LHC and the LEP colliders. As an example, Fig.~\ref{LHCSN} displays measurements of the seismic noise power spectrum in the LHC tunnel~\cite{CDGC}. Due to the large size of the ring, 26.7 km, even very small strains result in absolute deformations of the circumference exceeding 1 mm for large tides and large earthquakes occurring anywhere on the globe (magnitudes of 7 or higher). The beam position monitor system is currently the best tool to observe such deformations. Changes of the circumference are particularly well resolved with around 500 measurements along the LHC circumference. The resolutions on the average circumference measurement reaches the micrometer scale over time intervals of one hour. Strains (relative circumference changes) down to around $10^{-11}$ may be resolved.

\subsection{Ground Vibration at the SSRF Site\\ {\it R.~Deng}}

The mechanical stability of key components in an accelerator is one of the main contributions to the beam orbit stability. A series of measurements and analyses of ground vibration at the SSRF site have been carried out since 2004. Conventional facilities were located outside the main building so as to minimize the vibration generated by mechanical noise due to pumps, etc., and a long-term vibration monitoring system was built to monitor the ground vibration of SSRF. Recently, in view of the SXFEL and SHINE projects, more measurements are being performed for the shafts and tunnels about 30m underground, where the accelerator will be located, and a permanent vibration monitoring system is also considered. In addition, more investigation and optimization for the mechanical engineering are underway, along with the operation of the accelerator, to realize the beam stability goal for SSRF.

\subsection{Response of a 
storage-ring beam to a gravitational wave \\ {\it K.~Oide}}
\label{sec:Oide}

\begin{figure}[hbp]
\centering
\includegraphics[width=0.6\textwidth]{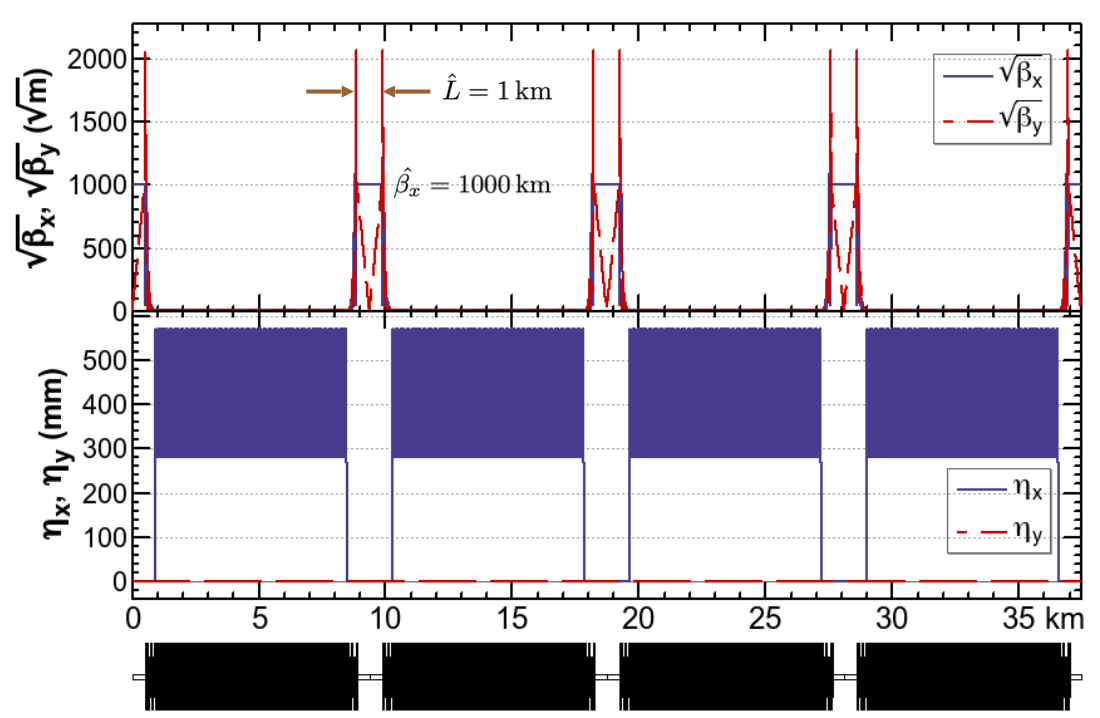}
\caption{A 37-km storage ring lattice optimized as a GW antenna.}
\label{KL}
\end{figure}

Summary of the presentation:

\begin{itemize}
\item  The beam in a storage ring can respond to a gravitational wave (GW), somewhat similarly to the Weber bar. A helpful reference is the book by Misner, Thorne, and Wheeler \cite{Misner:100593}.
\item The betatron motion of the beam in a ring can respond resonantly to the GW.
\item A special beam optics ``beam antenna'' may enhance the sensitivity. One example of a 
37-km storage ring lattice optimized as a GW antenna is shown in Fig.~\ref{KL}. 
\item There will be many noise sources to overcome: thermal motion of quadrupoles, beam fluctuation due to emittance, synchrotron radiation and the acceleration, etc.
\end{itemize}

Questions:
\begin{itemize}
\item Does the force depend on the choice of the coordinate system? --- A combination of motion of bunches to single out the ``quadrupole mode'' does not depend on the choice of the coordinates.
\item Are any sources beyond 10 kHz range conceivable in the universe?
\item This scheme may be suitable for GWs from the crab pulsar (60 Hz, continuous). If FCC is used, the necessary betatron tune is 0.02, at which operation is quite possible.
\end{itemize}

\subsection{Detection of millihertz gravitational waves using storage rings\\
{\it S.~Rao, M.~Br\"{u}ggen and J.~Liske}}
\label{sec:Rao}

Based on our study  \cite{PhysRevD.102.122006}, we showed that a storage ring can be turned into a gravitational wave (GW) observatory which is sensitive to millihertz GWs from astrophysical sources. This can be achieved, in principle, by making accurate timing measurements of a test mass circulating in a storage ring and, after resolving the noise sources, to observe a unique GW signal which causes the net travel time of the test mass to deviate from its nominal value (expected value in the absence of GWs), causing the change in travel time from its nominal value to oscillate with the periodicity of the GWs,  while also being modulated by the effect of Earth's rotation. The pros of this detection technique are that, it potentially allows for an earth-based gravitational wave observatory that is sensitve to millihertz GWs, which can also complement the space-based laser interferometer detectors (like LISA, planned to be launched in 2034), allowing for better triangulation of the position of millihertz GW sources in the sky ( i.e. improvement of GW source sky-localization), which is useful for electromagnetic follow-up studies of the GW event via astronomical telescopes. 

The effect of GWs studied in our work is thus along the longitudinal direction ( i.e. along the travel direction of the particle), and our calculations for this effect (excluding the influence of the Rf system) are parametrically in agreement with those by R.T.~D~Agnolo et al. in their works \cite{dagnolo1,dagnolo2}, cited with permission. Therefore, the general relativistic calculations for this GW effect done from two different reference frames (metric formalism and Riemann tensor formalism), give effectively the same results. Several previous works \cite{Dong:2002sr,ZerZion:1998tg,Ivanov:2002ng,vanHolten:1999yv,Faber:2017ovh} 
missed this effect, because they only considered the particular orientation of the GW propagating perpendicular to the ring plane, where this effect becomes null in the leading order of $h$ (the GW strain amplitude), and only the effect along the radial direction is maximised, which has an indirect contribution along the longitudinal direction with leading order of $h^2$, negligible compared to the effect we studied, since $h$ is a very small number. 

The open question is in figuring out a feasible realization of this experiment concept, and the conclusions of the SRGW2021 workshop (at least for the LHC and some other current CERN facilities), is that there is a technological limitation due to the sources of noise. In the presence of an RF system, the RF can potentially cause a damping of this GW effect (as shown by R.T. D'Agnolo et al.), or add its own noise. In the absence of an RF system (in a coasting beam, for instance), there are yet other sources of noise (such as collisions of the ions with with rest gas molecules etc., pointed out by Fritz Caspers). Hence, if we can technologically optimise the experiment setup and/or account for all noise sources, then we can potentially observe the GW effect using this detection technique. 
We, the authors of Ref.~\cite{PhysRevD.102.122006}, are currently working on numerical simulations of the expected GW signal, accounting for realistic GW waveforms from astrophysical sources, the effect of Earth's rotation etc. On another avenue, we are investigating whether single ion storage rings could be feasible candidates to conduct this experiment in the future.

\subsection{Storage rings as detectors for relic gravitational-wave
  background?\\
  {\it A.~N.~Ivanov, A.~P.~ Kobushkin, M.~Wellenzohn}}
\label{sec:Ivanov}

We argue that storage rings can be used for the detection of
low-frequency gravitational-wave background.  We explain the
systematic shrinkage of the machine circumference of the storage ring
of the SPring-8, observed by Takao and Shimada (Proceedings of EPAC
2000, Vienna, 2000, p.1572), by the influence of the relic
gravitational-wave background.  We show that the forces, related to
the {\it stiffness} of the physical structures, governing the path of
the beam, can be neglected for the analysis of the shrinkage of the
machine circumference caused by the relic gravitational-wave
background. We show the systematic shrinkage of the machine
circumference can be explained by a relic gravitational-wave
background even if it is treated as a stochastic system incoming on
the plane of the machine circumference from all quarters of the
Universe. We show that the rate of the systematic shrinkage of the
machine circumference does not depend on the radius of the storage
ring and it should be universal for storage rings with any radius.
Details can be found in Ref.~\cite{Ivanov:2002ng}.

\subsection{Gravitational Waves at Particle Storage Rings \\
{\it R. T.~D'Agnolo}}
\label{sec:Dagnolo}


\paragraph{Introduction}
Gravitational Waves (GWs) are an important probe of the Universe today and are a unique tool to shed light on its history before photon decoupling and Big Bang Nucleosynthesis~\cite{Schutz:1999xj, Cutler:2002me}. There are several accepted astrophysical sources of GWs such as white dwarf, neutron star and black hole binaries and a number of hypothetical cosmological sources including reheating after inflation and phase transitions in the early universe. The many orders of magnitude in frequency and amplitude that are still unexplored, strongly motivate the pursuit of new experimental concepts.

Here we discuss the possible detection of gravitational waves at circular colliders through measurements of the beam position and revolution period. In the following we use the LHC as a case study and then generalize to other storage rings. Most of the discussion is a review of textbook concepts in general relativity that help us build the right intuition for the LHC.

\paragraph{Basics of GW Detection}
A gravitational wave can have at least two effects on a storage rings: 1) It can interact with its electromagnetic fields; 2) it can displace the particles in the ring in the radial, vertical or longitudinal direction. The first possibility is discussed in another talk at this workshop (see Sebastian Ellis' contribution and~\cite{Pegoraro:1977uv}). I focus on the second possibility and in particular on the longitudinal direction, which is the most promising direction for detection, out of the three. 

A priori it might seem hard to disentangle the different effects of a GW and understand what we can measure. In principle everything in a storage ring interacts with the gravitational wave at the same time: magnets, tunnel, protons, electromagnetic fields, and we observers. However if we focus on a small frequency range, the majority of potential detectors are rigid and do not appreciably respond to the wave. 

Any system in equilibrium displaced by a small amount $\delta$ responds as a harmonic oscillator 
\be
\ddot \delta +\frac{\dot \delta}{\tau_s}+\omega_s^2 \delta =0\, .
\ee
In the proper detector frame (described in the next Section) we can approximate the effect of the wave as 
\be
\ddot \delta +\frac{\dot \delta}{\tau_s}+\omega_s^2 \delta =\omega_g^2 f(\omega_g, t)\, .
\ee
Hence any system responds as
\be
\tilde \delta(\omega)=\frac{\omega_g^2\tilde f(\omega_g, \omega)}{\omega_s^2-\omega^2+i \omega/\tau_s}\, ,
\ee
and any object with $\omega_s \gg \omega_g$ appears as rigid, i.e., it is not appreciably affected by the GW:
\be
\tilde \delta(\omega)\simeq \frac{\omega_g^2}{\omega_s^2}\tilde f(\omega_g, \omega) \ll \tilde f(\omega_g, \omega)\, .
\ee
For example, protons at the LHC are kept in equilibrium in the longitudinal direction by the RF system~\cite{LHCrf}:
\be
\ddot \delta_l +\frac{\dot \delta_l}{\tau_l}+\omega_l^2 \delta_l =0\, ,
\ee
where $\omega_l \simeq 60$~Hz. Protons reaching the RF cavity later than their nominal revolution period see a larger electric field and are accelerated more, and the opposite is true for protons arriving earlier than they should. 

If we focus on a GW that changes the protons' revolution period, we have the maximal response of the system when the effect is resonant: $\omega_g \simeq \omega_l$. Clearly, for GWs of this frequency the magnets or an observer appear as rigid in the proper detector frame, so we can focus only on the protons that respond to the wave maximally at this frequency.


\paragraph{An Estimate in the Proper Detector Frame}
General relativity has a large gauge symmetry. All possible choices of coordinates
\be
x^\mu\to x^{\mu\prime}(x^\mu)
\ee 
describe the same physics. We can use this to our advantage by choosing a reference frame where Newtonian intuition applies. 
This is known as the {\it proper detector frame}. In this frame: 1) coordinate distances coincide with proper distances for non-relativistic objects; 2) rigid objects are not appreciably deformed by the wave; 3) at zero GW frequency $\omega_g=0$ we recover flat space; 4) in the cases of interest to us the GW acts as a Newtonian force.

This is especially convenient in the case under consideration, because we want to study gravitational waves that do not oscillate appreciably over the size of the LHC: $L_{\rm LHC}/\lambda_g \ll 1$. In this regime we can expand the metric to second order in $L_{\rm LHC}/\lambda_g$. In the proper detector frame this gives
\be
ds^2=- dt^2 \left[1+R_{0i0j}x^ix^j\right]+2 dx^i dt \left(\frac{2}{3}R_{0ijk}x^j x^k\right)+dx^idx^j\left(\delta_{ij}-\frac{1}{3}R_{ikjl}x^k x^l\right)\, , \label{eq:metric}
\ee
where $R_{\mu\nu\rho\sigma}$ is the Riemann tensor and the spatial coordinates $x^i$ are measured from an origin chosen by the observer (for instance the center of the LHC ring). For a monochromatic wave of frequency $\omega_g$, neglecting  indexes and $\mathcal{O}(1)$ factors describing polarizations, we have the order of magnitude estimate
\be
R \sim \omega_g^2 h \cos(\omega_g t +\phi)\, ,
\ee
where $h$ is a tiny ($\sim 10^{-20}$) dimensionless number. For a derivation of the above results, including the metric in the detector frame, we refer to the first chapter of~\cite{Maggiore:1900zz}. We have neglected the Earth's gravity, which contributes to the $00$ component of the metric, but does not affect appreciably the motion of the protons that are kept on a stable circular orbit by the LHC magnets and RF system.

To compute the effect of the wave described by the metric in Eq.~\eqref{eq:metric} we imagine an observer fixed at a specific position along the ring. The observer moves along a geodesic with coordinates $x^\mu(\tau)$, where $\tau$ is the proper time. The protons move along the ring and are in general on a different geodesic $y^\mu(\tau)=x^\mu(\tau)+\xi^\mu(\tau)$. For small displacements $\xi^\mu$ compared to the wavelength of the GW ($L_{\rm LHC}/\lambda_g \ll 1$) we can use the equation for geodesic deviation~\cite{Maggiore:1900zz}
\be
\frac{D^2 \xi^\mu}{D\tau^2}=- R^\mu_{\nu\rho\sigma}\xi^\rho \frac{d x^\nu}{d\tau}\frac{d x^\sigma}{d\tau}\, .
\ee
In the proper detector frame this reduces to
\be
\frac{d^2 \xi^i}{d\tau^2}=-R^i_{0j0}\xi^j \left(\frac{dt}{d\tau}\right)^2 + ...
\ee
The terms represented by the ellipses are in the form $R^i_{0jl}\xi^j v^l$ and $R^i_{kjl}\xi^j v^l v^k$. So for highly relativistic protons, $v\simeq 1$, they are of the same order as $R^i_{0j0}\xi^j$ (as is the case at the LHC), but can be neglected for non-relativistic particles. 
Since we are interested in an order of magnitude estimate we approximate all the terms on the right-hand side of the equation with their parametric value at the LHC for a monochromatic wave propagating in the plane of the accelerator: $R^i_{0j0}\xi^j, R^i_{0jl}\xi^j v^l, R^i_{kjl}\xi^j v^l v^k \simeq \omega_g^2 h L_{\rm LHC} \cos(\omega_g t+\phi)$. 

To get the equation of motion for the protons in our reference frame we notice that for a non-relativistic observer $dt=d\tau+\mathcal{O}(h)$, so
\be
\frac{d^2 \xi^i}{dt^2}=-R^i_{0j0}\xi^j + ... \, ,
\ee
where the $\xi^i$ are the coordinates of the protons in a reference frame moving along the geodesic of the observer and $\frac{d^2 \xi^i}{dt^2}$ is their acceleration. Note that we would have obtained the same result in the reference frame of the protons $dt=\gamma d\tau+\mathcal{O}(h)$.

The previous equation describes the effect of the wave as a Newtonian force. To describe correctly the motion of the protons in this frame we just need to add to the previous equation any other external forces. In particular, if we are measuring variations over the revolution period of the protons we have
\be
\frac{d^2 \delta\xi^l}{dt^2}+\frac{1}{\tau_l}\frac{d\delta\xi^l}{dt}+\omega_l^2 \delta\xi^l\simeq \omega_g^2 h L_{\rm LHC} \cos(\omega_g t+\phi)\, , \label{eq:motion}
\ee 
where $l$ is the longitudinal direction along the circumference. We could easily have obtained the same result from dimensional analysis.

\paragraph{Possible Avenues towards Detection}
If we solve Eq.~\eqref{eq:motion} on resonance to maximize the effect, we obtain
\be
\frac{\Delta T}{T} \simeq h \omega_l \tau_l\, .
\ee
At the LHC we have $\frac{\Delta T}{T} \simeq 10^{-7}$ from the phase measurement in the RF system~\cite{LHCrf} and $\omega_l \simeq 60$~Hz. We thus get a sensitivity (without backgrounds) 
\be
h \gtrsim 10^{-13} \left(\frac{2 \pi\times 10\; {\rm Hz}}{\omega_l}\right)\left(\frac{10\;{\rm hours}}{\tau_l}\right)\left(\frac{\Delta T/T}{10^{-7}}\right)\, .
\ee
This is about 7 orders of magnitude larger than what we expect from realistic GW sources~\cite{Maggiore:1999vm}. However, there is in principle a path towards vastly increasing the sensitivity of a storage ring:
\begin{enumerate}
\item The effect saturates when $T\simeq \omega_g^{-1}$. So we can either increase $\omega_l$ or increase $T$ by about three orders of magnitude compared to the LHC, where $T\simeq 1/(11\;{\rm kHz})$, increasing the sensitivity on $h$ by the same amount. Increasing $\omega_l$ reduces the amplitude of expected astrophysical signals, so it is better to consider rings with either non-relativistic protons or sizes even larger than the LHC.
\item One can improve the time resolution of the measurement. Using electro-optical sampling as in other storage rings~\cite{Casalbuoni:2008zz} could improve $\Delta T/T$ by one order of magnitude. 
\item Decreasing the energy of the protons also goes in the direction of increasing $\tau_l$. Maybe one can envision operating a storage ring in stable conditions for days or even weeks.
\end{enumerate}
The three above points indicate future avenues of investigation that might lead to the detection of GWs at storage rings. Another interesting option is to make the protons as freely-falling as possible, i.e., $\omega_l \to 0$. In this limit the wave can act coherently on $T$ for a very long time $1/\omega_g \simeq 1/\omega_l$. A calculation neglecting the RF system was presented in another contribution to this workshop (see the talk by Suvrat Rao). This suggests that in this regime mHz GWs might be detected at a storage ring. However the problem of keeping the protons in a stable orbit in these conditions remains open at the time of writing this article.

In conclusion, from the very simple estimates presented here, measuring GWs using particle motion inside storage rings is quite challenging. However, I consider the existence of a conceptual path towards GW-level sensitivity, as outlined in this Section, interesting and deserving of further study.

\subsection{Gravitational Synchrotron Radiation from Storage Rings \\
{\it Pisin Chen}}
\label{sec:Chen}

In this presentation, 
we show that relativistic charged particles executing circular orbital motion, as in a storage ring, can emit gravitational waves through two channels. One is the gravitational synchrotron radiation (GSR) emitted directly by the massive particle; the other is ‘resonant conversion’, i.e., the Gertsenshtein effect, which, in this case, converts the electromagnetic synchrotron radiation (EMSR) to GWs. 
It is shown that the dominant frequency of the direct GSR is its fundamental mode, i.e., 
$\omega_0=c/\rho$, where $\rho$ is the radius of the storage ring. In the case of CERN LHC, $\omega_0 \sim 10^{4}-10^{5}$~Hz.
The dominant frequency of resonant EMSR-GSR conversion is a factor $\gamma^3$ higher, and for LHC it is around 
$\omega_c=\gamma^3 \omega_0\sim 10^{13}$~Hz, with wavelength at $\sim \mu$m, which, if detected, would be the first observation of gravitons.         
Since the observation point can be located near the storage ring, we find that the spacetime metric perturbation can be 
$h_{\rm GSR}  \sim  n_b N \times 10^{-31}\sim 10^{-18}$ for the LHC. 
More details can be found in Refs.~\cite{Chen:1994vk,Chen:1994ch}.

Open questions and/or planned next steps: 
The deduced scaling law for the spacetime perturbation, 
$h\sim (\gamma^2 m_p^2 M_{\rm Pl}^2) (\rho/R)$, requires further confirmation.
Under what conditions can the bunch train be treated as radiating GSR coherently?
Is there a near-field effect in GSR and, if so, is that more favorable in terms of detection?
Would the heavy ion mode in LHC be preferable for GSR production? 

Additional Comments: 
The author would be most eager to collaborate with colleagues to push forward an experiment at CERN to generate and detect the direct GSR and the resonant EMSR-GSR conversion signals.

\subsection{On Gravitational Synchrotron Radiation\\ {\it J.~M.~Jowett}}
\label{sec:Jowett}

Gravitational synchrotron radiation from the future LEP, LHC and SSC beams were discussed among a number of people at CERN in the late 1980s, shortly before LEP started operation.  It was realised that these beams would be among the most powerful terrestrial sources of gravitational radiation, although the total radiated power would still only be of order $10^{-25}\, {\rm W}$.  

This presentation revisited the paper \cite{diambrini} which treated the subject in parallel to the familiar treatment of incoherent synchrotron radiation.    The numerical results of \cite{diambrini} were reproduced and extended to the beams in the HL-LHC and FCC-hh.   

The question had been raised whether the heavy-ion beams in hadron colliders could be a more powerful source.  
It was shown that a Pb nucleus radiates 167 times more power than a proton in the same magnetic field, the same $Z^6/m^4$ scaling as for photons.   However the total power is smaller because the number of Pb ions in HL-LHC or FCC-hh is about a factor 1000 smaller.     

The relation between these results, based on coherence conditions from \cite{diambrini}, and \cite{Chen:1994vk}, which considered that all particles in the ring radiate coherently, requires some clarification.

\subsection{Revisiting Gravitational Wave Detection in an SRF Cavity\\
{\it A.~Berlin, R.~T.~D'Agnolo, S.~A.~R.~Ellis}}
\label{sec:SEllis}

Since the discovery of gravitational waves (GWs) by the LIGO collaboration~\cite{Abbott:2016blz}, there has been much renewed interest in approaches to their detection (see, for example, the recent proposals~\cite{Graham:2012sy,Bertoldi:2019tck,Badurina:2019hst,Dimopoulos:2008sv,Canuel:2019abg,Aggarwal:2020umq,Aggarwal:2020olq,Herman:2020wao}). Many of these proposals seek to detect low-frequency (sub-10 Hz) GWs, where standard astrophysical sources are expected aplenty. However, less focus has been placed on high-frequency GWs, in the kHz$-$MHz range. The reasons for this include the expectation that no black holes of stellar origin can produce such high-frequency signals, and that the strength of the signal decreases significantly at higher frequency, making detection ever more difficult~\cite{Maggiore:1999vm,Aggarwal:2020olq}. However, proposals have existed for many years to detect high-frequency GWs~\cite{Pegoraro:1977uv,Pegoraro:1978gv,Caves:1979kq,Reece:1984gv}, and there has recently been a renewed interest in new techniques, driven by advances in quantum sensing technology, and by suggestions of potential sources~\cite{Caprini:2018mtu,Aggarwal:2020umq,Aggarwal:2020olq}.

In the late 1970s, proposals to search for GWs using superconducting radio-frequency (SRF) cavities was made by Pegoraro, Picasso and Radicati (PPR)~\cite{Pegoraro:1977uv,Pegoraro:1978gv} and by Caves~\cite{Caves:1979kq}. Their idea was that a cavity would be prepared with a mode split into two nearly-degenerate modes in a symmetric and anti-symmetric configuration, and one of the two modes would be pumped, while the other would be kept quiet. Gravitational waves passing through the detector would deform the cavity walls, thereby transferring power from the pumped mode into the quiet mode if the GW frequency $\w_G$ was matched to the splitting of the two cavity modes, $\w_G \simeq |\w_a - \w_s|$, where the subscripts denote the symmetric ($s$) and anti-symmetric ($a$) modes. On this basis, prototypes were constructed in the 1980s and early 2000s, culminating in the MAGO collaboration~\cite{Reece:1984gv,Bernard:2002ci,Ballantini:2003nt,Bernard:2001kp,Ballantini:2005am}. The approach was abandoned, however, due to political decisions prioritizing other GW searches where known astrophysical sources exist~\cite{Gemme}.

Due to the renewed theoretical and experimental interest in high-frequency GWs, and advances in SRF technology, we have revisited the use of carefully-controlled SRF cavities to detect gravitational waves. Similar to the PPR approach, the idea would be to design a cavity such that a low-order mode is split into a symmetric and anti-symmetric mode, to pump one of the two, and look for a signal in the quiet mode.
However, contrary to the PPR approach, we are interested in detecting GWs through the inverse Gertsenshtein effect (IGE)~\cite{Gertsenshtein}, whereby GWs convert to photons in a background electromagnetic field. The reason for considering this signal, rather than the mechanical one of PPR, is that mechanical vibrations can be a significant noise source at low mode separations $|\w_a - \w_s| \ll \w_{a,s}$~\cite{Berlin:2019ahk,Berlin:2020vrk}, as well as being a transducer of a possible signal. Because of the potential for noise vibrations to cause significant cavity frequency drift, we propose to use a precise active feedback system to control the cavity frequency drift to be smaller than the cavity bandwidth, $\Delta\w \sim \w/Q$. This active feedback would not only damp mechanical noise, but also any mechanically-induced GW signal, thereby necessitating the study of the IGE signal from GWs. Recently, the DarkSRF collaboration at FNAL has achieved $\Delta\omega \sim 0.1$~Hz in a cavity operating at $\w \sim$~GHz, providing the proof of concept that such control can be achieved with current SRF technology~\cite{DarkSRF}.

Employing the approach of searching for the GWs converting the pumped mode into the quiet mode via the IGE we expect, based on preliminary calculations, that the strain sensitivity should be on the order of $h\sim10^{-22}$ for a thermal noise-limited device operating at $\omega \sim 100$~ MHz. The exact details of the computation will be found in a forthcoming publication~\cite{GravWaves}. 

Future directions of this research include establishing an experimental collaboration to build and operate the SRF cavity, as well as conducting theoretical work on potential sources at high frequencies. Experimental activities might be combined with a similar approach to the search for axion dark matter, which employs similar SRF technology~\cite{Berlin:2019ahk,Berlin:2020vrk}.

\subsection{An atom interferometer to search for ultralight dark matter and gravitational waves at CERN?\\
{\it O.~Buchmueller and J.~R.~Ellis}} 
\label{sec:Buchmueller}

We outline the scientific opportunities of a proposed programme using ground- and space-based Atom Interferometry experiments   
to search for ultra-light dark matter, to explore gravitational waves in the mid-frequency 
range between the peak sensitivities of the LISA and LIGO/Virgo/ KAGRA/INDIGO/Einstein Telescope/Cosmic Explorer
experiments, and to probe other frontiers in fundamental physics. This programme would complement other planned
searches for dark matter, as well as probe mergers involving intermediate-mass black holes and explore early-universe cosmology.

\paragraph{The AION Project} 
The Atom Interferometer Observatory and Network (AION) is a project to explore ultra-light dark matter (ULDM)
and mid-frequency gravitational waves (GWs)~\cite{Badurina:2019hst} that has recently received initial funding of about £9.6M
from the UK Quantum Technology for Fundamental Physics Programme and other sources. It is being
executed in national partnership with the UK National Quantum Technology Hub in Sensors and Timing in Birmingham 
and in international partnership with the MAGIS Collaboration and the Fermi National Laboratory in the US~\cite{Graham:2017pmn}.

\begin{figure}[hbp]
\centering
\includegraphics[width=0.8\textwidth]{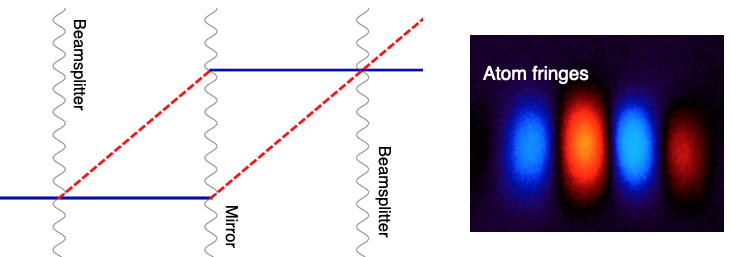}
\caption{The principle of atom interferometry~\cite{Arvanitaki:2016fyj}: 
blue lines are atoms in the ground state, red dashed lines are atoms in the
excited state and the wavy lines are laser pulses; and a typical interference pattern.}
\label{AI}
\end{figure}

The basic principle of an atom interferometer is illustrated in Fig.~\ref{AI}~\cite{Arvanitaki:2016fyj}. A laser splits a cloud
of atoms into ground and excited states that follow different trajectories because of the momentum
kicks they have undergone. After following different space-time paths, they exhibit an interference
pattern upon recombination. The same laser beam strikes two or more clouds, and waves of ULDM or
passing GWs can generate differences between their interference patterns.

Stage 1 of AION is to build and commission a 10m detector using atom interferometry, developing
existing technology and exploring possible infrastructure for a 100m that would be Stage 2. 
In addition to building, commissioning and exploiting the 100m detector, Stage 2 would also
carry out a design study for a follow-on km-scale detector, 
which is planned as Stage 3. 
AION was selected in 2018 by STFC as a high-priority medium-scale project,
and will work in partnership with MAGIS in the US~\cite{Graham:2017pmn} to form a ``LIGO/Virgo-style''  network
and collaboration. Stage 1 is to be located in the Oxford University Physics Department,
while Stage 2 could 
be placed either at a UK national facility in Boulby or Daresbury, 
or possibly at CERN. 
Stage 4 of AION is a long-term objective to orbit a pair of satellite detectors
in medium Earth orbit called AEDGE~\cite{AEDGE}, which was proposed to ESA in response to its Voyage2050 call
for possible mission concepts.
Members of AION took the lead in proposing AEDGE and developing its science case, 
bringing together collaborators  from European and Chinese groups including those working on the
MIGA~\cite{MIGA}, MAGIA~\cite{MAGIA}, ELGAR~\cite{Canuel:2019abg} and ZAIGA~\cite{Zhan:2019quq} projects.
Stages 3 and 4 will probably both require funding on an international level via the EU, ESA, etc., 
and the AION collaboration has already started to build the foundation for such an international effort. 

AION Stage 1 will be built on the basement level of the new purpose-built building
of the Oxford Physics Department, which has 14.7m headroom. The site has stable concrete construction
and space for a temperature-controlled laser laboratory. In addition to benefiting from this
world-class infrastructure, AION will receive strong engineering support from the nearby
STFC Rutherford-Appleton Laboratory (RAL). The focus during the first 30 months of the project
will be on establishing the prerequisites for the 10m detector, including a network of
ultracold strontium atom sources and laser laboratories in Birmingham, Cambridge, Imperial College, Oxford and RAL,
and training early-career scientists in the necessary expertise.
These laboratories are expected to be fully operational in late 2021.
In parallel, the final design for 10m detector will be developed, and preparations made for its
construction and subsequent physics exploitation. As already mentioned, this development phase will be in partnership
with the MAGIS experiment in the US, which is to be installed in a vertical access shaft for the
Fermilab neutrino beam. Several of the other atom interferometer projects, such as the MIGA and ELGAR projects in Europe and the ZAIGA project in China, are being prepared on similar
timescales.

\begin{figure}[htb]
\centering
\includegraphics[width=0.4\textwidth]{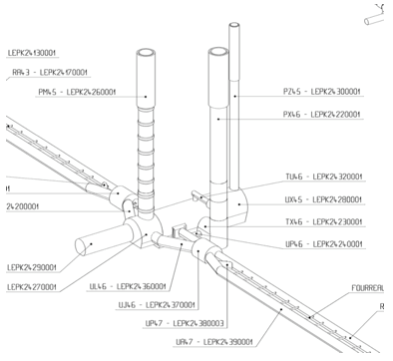}
\caption{One possible location of the 100m Stage 2 of AION at CERN is in the PX46 access shaft to the LHC.}
\label{CERN}
\end{figure}

Possible locations for the 100m Stage 2 of AION include a tower at the STFC Daresbury Laboratory
and a shaft in the STFC Boulby underground laboratory. A third possibility is one of the LHC
access shafts, namely PX46, as illustrated in Fig.~\ref{CERN}. 
Its height (143m) and diameter (10.1m) are ideal for AION, and the
technical infrastructure available at CERN is very attractive. First radiation studies have also 
been quite promising, but more work is needed to determine if PX46 could be a valid option for
Stage 2 of AION. We are currently working with the CERN Physics Beyond Colliders team on a
feasibility study, in particular for the arrangement of shielding that would be required if the LHC
were to suffer a catastrophic beam loss near Point 4 while people are working in PX46.

\begin{figure}[hbp]
\centering
\includegraphics[width=0.9\textwidth]{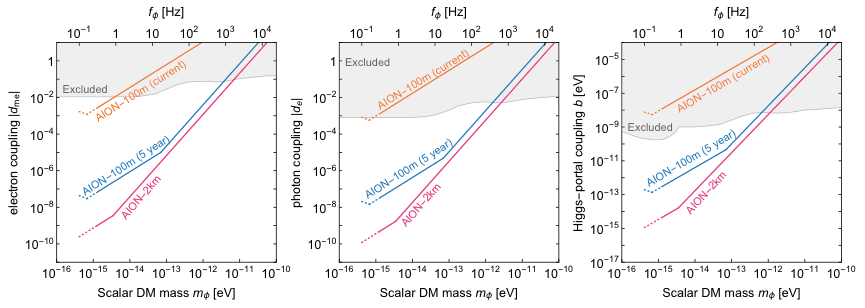}
\caption{The sensitivities of various AION stages to ULDM couplings to electrons (left), photons (middle)
and through a Higgs portal (right)~\cite{Badurina:2019hst}}
\label{DM}
\end{figure}

\begin{figure}[hbp]
\centering
\includegraphics[width=0.6\textwidth]{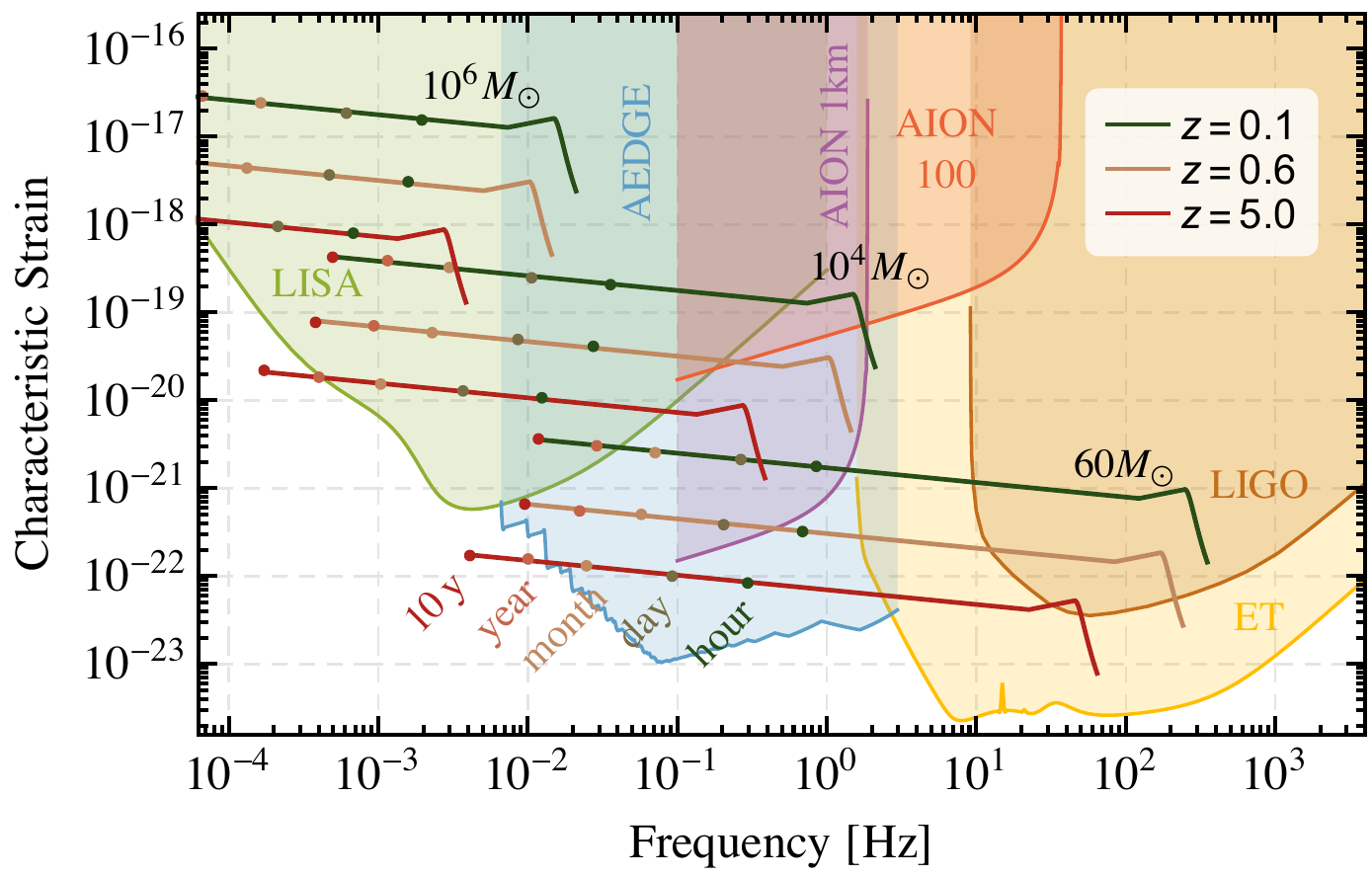}
\caption{The sensitivities of various AION stages to gravitational wave strain as functions of frequency,
compared with the sensitivities of LIGO, LISA and ET, indicating also the signals expected for mergers of black holes
of various masses at different redshifts $z$~\cite{AEDGE}.}
\label{h}
\end{figure}

\paragraph{Science Objectives} 
The main physics goal of Stage 1 will be the search for ULDM in the form of some scalar
field with a mass around $10^{-15} - 10^{-14}$~eV, exploring new domains of parameter space 
and complementing other searches, as illustrated in Fig.~\ref{DM}~\cite{Badurina:2019hst}. 
Possible extensions to pseudoscalar and vector dark matter fields
are also under study. The search for ULDM will continue with subsequent stages of AION, and will
be accompanied by growing sensitivities to GWs in the mid-frequency band between the peak
sensitivities of LISA (at $\sim 10^{-2}$~Hz) and LIGO, Virgo, KAGRA and the proposed Einstein Telescope (ET)
(at $\sim 10^{2}$~Hz), as illustrated in Fig.~\ref{h}~\cite{AEDGE}. 
Prime targets as sources of mid-frequency GWs include the mergers of
black holes, cosmological phase transitions and cosmic strings.

\begin{figure}[htbp]
\centering
\includegraphics[width=0.35\textwidth]{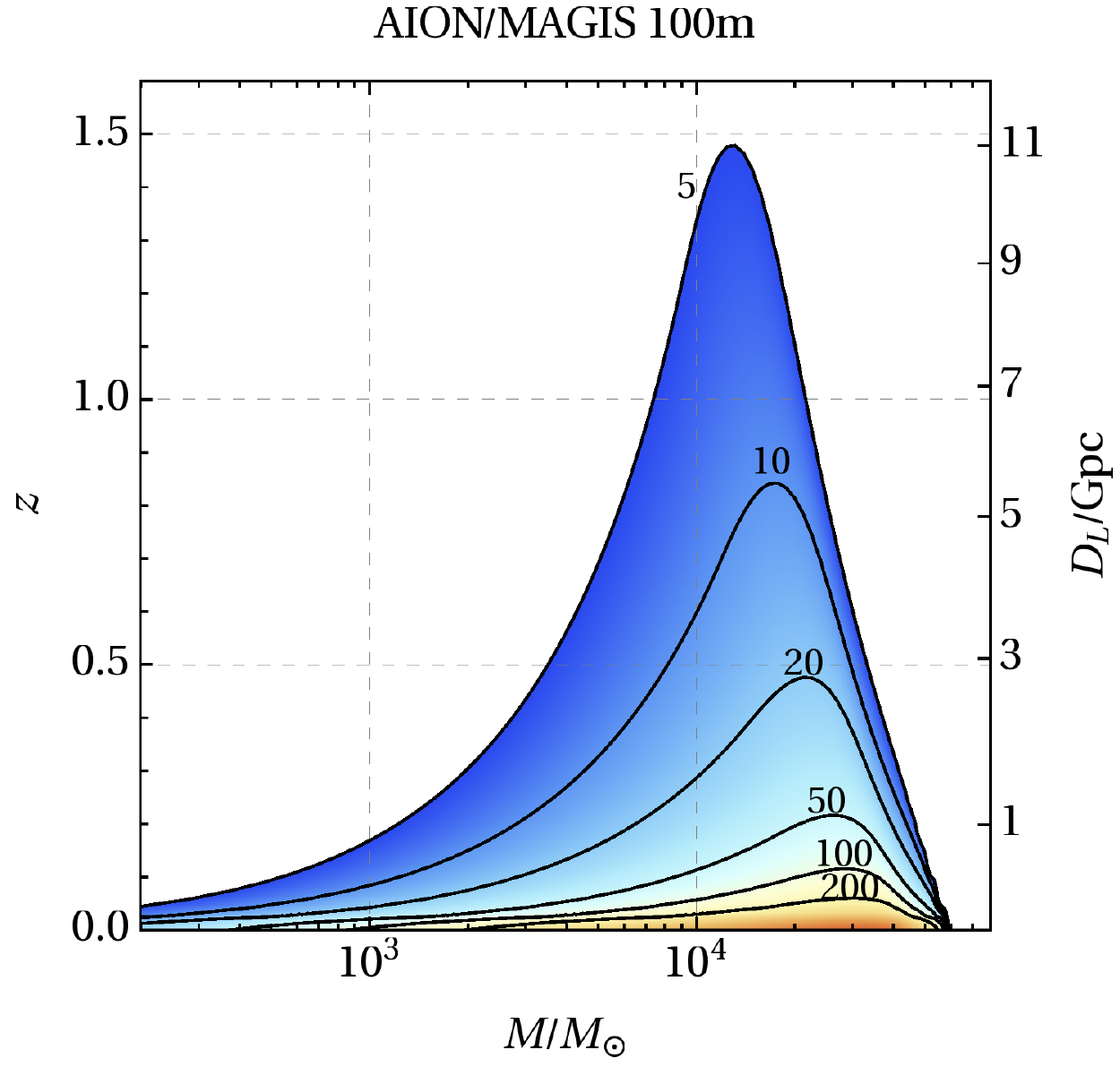}
\caption{The estimated signal-to-noise ratios attainable with AION 100m for mergers of
black holes of masses $M$ at different redshifts  $z$.~\cite{Badurina:2019hst}}
\label{Matterhorn}
\end{figure}

High-priority astrophysical questions to be investigated by studies of GWs in the mid-frequency
band include how supermassive black holes in galactic centres were formed, e.g., by mergers of
intermediate-mass black holes. The capabilities of AION 100m for observing such mergers are shown
in Fig.~\ref{Matterhorn}. There will also be synergies with LISA and LIGO/Virgo/KAGRA/ET
observations through observations over long periods that make possible early warnings and localizations
of future mergers. These will also make possible sensitive probes of fundamental physics, such as
the possible mass of the graviton and violation of Lorentz invariance by GWs~\cite{EV}.

Other topics beyond the Standard Model (SM) of particle physics include the possible observation
of GWs generated during a first-order phase transition during the early Universe. 
Among the extensions of the SM that AION could study
are additional $H^6/\Lambda^2$ interactions and extra $U(1)_{B-L}$ gauge bosons~\cite{Badurina:2019hst}. These scenarios
also have prospective collider signatures that are complementary to those from GWs.
AION could also search for evidence of a network of cosmic strings~\cite{EL}, as has recently been hinted
by the NANOGrav pulsar timing array~\cite{Arzoumanian:2020vkk}.

Other possible topics in fundamental physics include high-precision measurements of the Newton constant and the 
gravitational redshift, probes of Bell inequalities, the equivalence principle and
 the constancy of fundamental parameters, and searching for fundamental
(i.e., not environmental) decoherence. Installing Stage 2 of AION at CERN would add many
novel features to CERN's programme in Physics Beyond Colliders.

\section{Final Workshop Discussion\\
{\it Moderated by J.~R.~Ellis} }

The final discussion centred on the following five main topics:
\begin{enumerate}
\item The proposal by K.~Oide (Section~\ref{sec:Oide}) of GW detection by resonant betatron oscillations in a storage ring, 
sensitive in the 10 kHz range;  
\item  A variant of the proposal by S.~Rao (Section~\ref{sec:Rao}) and R.~T.~D'Agnolo (Section~\ref{sec:Dagnolo}) of GW detection through the change in revolution period, 
but using a ``low-energy'' coasting ion beam without RF, aiming at sensitivity down to $10^{-5}$~Hz;
\item The proposal by P.~Chen (Section~\ref{sec:Chen}) (discussed also by J.~Jowett (Section~\ref{sec:Jowett}) of gravitational synchrotron radiation generated by the beam at frequencies close to the orbital frequency, $\sim 10^{4}$~Hz for LHC. Section~\ref{sec:Chen} also discussed the Gertsenshtein  effect, whereby  electromagnetic  synchrotron  radiation is converted to  GWs of much higher frequency;
\item The suggestion by S.~Ellis and collaborators (Section~\ref{sec:SEllis}) of heterodyne GW detection using SRF, which could be sensitive up to $\sim 10^{7}$~Hz;
\item 
The possibility suggested by O.~Buchmueller and J.~Ellis (Section~\ref{sec:Buchmueller}) of using an LHC access shaft to house a 100m atom interferometer targeting the
1 to $10^{-2}$~Hz range.
\end{enumerate} 

\begin{figure}[htbp]
\centering
\includegraphics[width=0.6\textwidth]{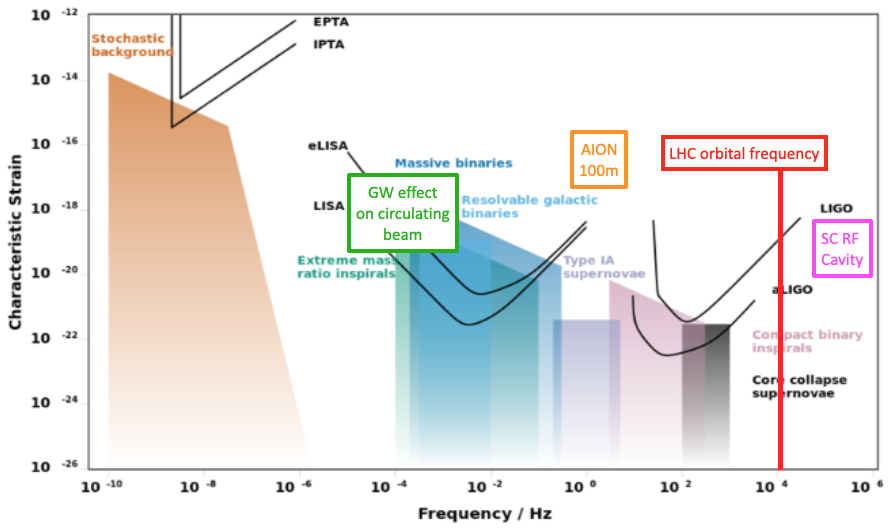}
\caption{The landscape of the GW frequency spectrum, illustrating the ranges in which various astrophysical and cosmological
sources are expected, and the sensitivities of some present and planned experiments. Also shown are the frequency ranges
targeted by some of the proposals discussed during this Workshop.}
\label{FrequencyPlot}
\end{figure}

Figure~\ref{FrequencyPlot} illustrates the ways in which these suggestions might complement the frequency coverage of existing,
planned and proposed GW experiments. We now review some of the points that emerged during discussions of these suggestions.

Proposal~1 is aimed at transverse beam perturbations at frequencies around the orbital frequency.
Key issues include the design of the storage ring, the suppression
of noise sources and the development of beam position monitors with nanometre precision that do not degrade the beam characteristics
through impedance effects. It was suggested in the 
Discussion that the CERN AD and ELENA facilities
could be used to study possible noise sources.

Similar issues arise in connection with Proposal~2 for studying longitudinal perturbations of the beam, which might be able to access part of the gap in frequencies $\lesssim 10^{-3}$~Hz where
LISA loses some sensitivity.
However, there are issues concerning the long-term stability of a
beam coasting in the absence of RF and the development of picosecond time resolution. The suggestion of using a low-energy ion beam emerged during the Workshop, and will be explored
by CERN accelerator physicists as well as by the authors of Ref.~\cite{PhysRevD.102.122006}. 

Concerning the generation of gravitational synchrotron radiation by a
circulating beam (Proposal~3), there is not yet a consensus on the strength of this GW source, 
which is likely to be very small but merits urgent clarification. It
would peak at frequencies above those where the sensitivities of LIGO and 
similar laser interferometers are reduced and astrophysical sources may be weaker.

Proposal~4 for the detection of high-frequency GWs using an SRF cavity could benefit from synergies
with the suggestion~\cite{Berlin:2019ahk,Berlin:2020vrk} to use similar cavities to search for axions, which is currently being
discussed at SLAC and CERN. 

Proposal~5 is aimed at the frequency gap between LIGO/Virgo at $\gtrsim 100$~Hz and LISA at 
$\lesssim 10^{-2}$~Hz. The requisite atom interferometer technology is being developed by the MAGIS Collaboration in the US 
as well as by the AION Collaboration in the UK and other groups in Europe and China. The AION Collaboration has chosen the Oxford
Physics Department as the site for its precursor 10m experiment, and is currently considering one of the LHC access shafts for its
100m successor. The possibility of being hosted by CERN's excellent infrastructure is very attractive, though other considerations may affect the choice of site. The radiation protection issue at CERN is under study, and it would
be interesting to extend the LHC seismic measurements shown in Fig.~\ref{LHCSN} to the LHC access shaft of interest.

One interesting aspect common to Proposals 4 and 5 is that they could/would be located close to the LHC beam pipe, where they 
could benefit from any near-field enhancement of a gravitational synchrotron radiation signal generated by the beam. 
For the same reason, the LHC underground
areas might also be interesting locations for other experiments searching for high-frequency GWs~\cite{Aggarwal:2020olq}.

We recall in this connection that the classical gravitational perturbation generated by a 
relativistic particle source is well understood~\cite{AS}, 
and can be applied to bunches of particles in a storage ring, as discussed recently in~\cite{SRB}. This reference
considers various possible detection schemes, including a monolithic pendulum, and it would be interesting to extend these
studies to other detectors that could be located close to the LHC beams, 
such as Proposals 4 and 5 and those reviewed in~\cite{Aggarwal:2020olq}.  

Many of the Proposals discussed during this Workshop involve elements of speculation, are at relatively early stages
of development, and require considerable conceptual and technical development. That said, the prospects of using a storage
ring as a GW detector or source are fascinating, and the opportunities for using CERN's technical infrastructure to add GW
physics to its Physics Beyond Colliders programme~\cite{PBC} are too interesting to be ignored.

\section*{Acknowledgements}
This work was 
supported in part by the European Commission under 
the HORIZON 2020 project ARIES no.~730871. The work of O.B. and J.E. was supported in part by STFC 
grant ST/T00679X/1 to the AION project, and the work of J.E. was also
supported in part by STFC grant ST/T000759/1 and by Estonian Research Council grant MOBTT5.
The work of S.R, M.B. and J.L. was was supported by the Deutsche Forschungsgemeinschaft 
(DFG, German Research Foundation) under Germany’s 
Excellence Strategy—EXC 2121 “Quantum Universe”—390833306.

\newpage
\bibliography{bibliography}

\appendix
\section{Workshop Programme and Schedule}
As coordinators of ARIES WP6 and its task 6.6, 
Giuliano Franchetti, Marco Zanetti and Frank Zimmermann 
jointly co-chaired this workshop.  
The SRGW2021 programme had been 
established with the help of an 
International Scientific Programme Committee consisting of
William A.~Barletta (MIT), Pisin Chen (NTU), 
Raffaele~Tito D'Agnolo (IPhT), Raffaele Flaminio (LAPP), 
Giuliano Franchetti (GSI), Shyh-Yuan Lee (Indiana U.),  
Katsunobu Oide (CERN \& KEK), Qing Qin (ESRF \& U.~Peking)
J\"{o}rg Wenninger (CERN), Marco Zanetti (U. Padova) and
Frank Zimmermann (CERN). 
The detailed SRGW2021 workshop schedule with sessions, presentations 
and speakers is presented in Table \ref{tab:schedule}.

\setlength{\fboxsep}{4pt}
\tabulinesep=1.2mm
\setlength{\tabcolsep}{4pt}

\begin{table}[htb]
\caption{SRGW2021 workshop sessions, 
session chairs, talks, and speakers}
\label{tab:schedule} 
\begin{center} 
\hspace*{-5 mm}
\begin{tabular}{lll}
\hline\hline
{\bf Session 1}
& 
{\bf Introduction to Gravitational Waves and}
&
{\bf Pisin Chen, NTU (Chair)}\\
{\bf 
2/2/21} 
& 
{\bf their Effects }
& 
{\bf 
 }
\\
& 
Welcome and Introduction 
& 
Frank Zimmermann, CERN
\\
& {A Brief History of Gravitational Waves} &
{Jorge Cervantes Cota, ININ, Mexico
}
\\
& 
{Expected Sources of Gravitational Waves}
& 
{Bangalore Sathyaprakash, 
Pennsylvania} \\ & &  ~~~State U., USA, 
\& Cardiff U., UK 
\\
& 
Response of storage-ring beam to  gravitational 
& Katsunobu Oide, KEK \& CERN 

\\ 
&  
~~~waves - preliminary considerations
& \\
& Discussion & \\
\hline
{\bf Session 2}
& 
{\bf Measurements and sensitivity }
&
{\bf Shyh-Yuan Lee, Indiana U. 
 (Chair)}\\
{\bf 
2/18/21} 
& 
& 
{\bf 
 }
\\
& Measurement approach and sensitivities of & 
Raffaele Flaminio, IN2P3 LAPP\\ 
& 
~~~detectors like LIGO and VIRGO & \\
& 
Storage ring sensitivity to tides \& large-scale 
& J\"{o}rg Wenninger, CERN \\
& ~~~perturbations
- examples from LEP \& LHC
& 
\\
& Discussion & \\
\hline 
{\bf Session 3}
& 
{\bf Proposals and Schemes }
&
{\bf J\"{o}rg Wenninger, CERN (Chair)}\\
{\bf 3/4/21 } & & \\
& 
Detection of gravitational waves in circular  
& 
Suvrat Rao, Hamburg U.\\
& 
~~~particle accelerators --- a proposal for the LHC & 
\\
& 
Storage rings as detectors for relic & 
Andrei Ivanov, Vienna U. 
\\
& 
~~~gravitational-wave background? &
\\ &
Update on Theoretical Effects of Gravitational 
& Raffaele Tito D'Agnolo, IPhT
\\ 
& 
~~~Waves on Storage Rings
& \\
& 
Discussion& \\
\hline
{\bf Session 4 } & 
{\bf GW generation \& detection}
& {\bf Frank Zimmermann, CERN (Chair)} \\
{\bf 3/11/21} & & \\
& 
Radiofrequency cavities \& gravitational 
& 
Sebastian Ellis, IPhT \\
& wave signals?  & \\
& Using storage rings as a GW source & 
Pisin Chen, NTU \\
& GSR, some history revisited, and FCC-hh & 
John Jowett, GSI \\
& 
Use of atom-interferometry for possible GW   & 
Oliver Buchmueller, ICL \\
&  ~~~detection \& other gravity experiments & \& John Ellis, KCL \\
\\
& Discussion & \\
\hline
{\bf Session 5 } & 
{\bf Summary and Outlook}
& {\bf Giuliano Franchetti, GSI (Chair) }\\
{\bf 3/18/21} & & \\
& Ground Vibration at SSRF Site & Rongbing Deng, SARI Shanghai \\
& 
{\it Contributed:} What Governs Flow of Energy? & Peter Cameron, BNL (ret.) \\
& ~~~Questions on Grav.~Impedance Matching  & \\
& Concluding Discussion & Moderated by John~Ellis, KCL

\\
\hline\hline
\end{tabular} 
\end{center}
\end{table}

\end{document}